\documentclass[12pt]{article}
\usepackage{amssymb}
\usepackage{amsmath}
\usepackage{amsfonts}

\begin{document}

\author{C. Bizdadea\thanks{%
e-mail address: bizdadea@central.ucv.ro}, D. Cornea\thanks{%
e-mail address: dcornea@central.ucv.ro}, S. O. Saliu\thanks{%
e-mail address: osaliu@central.ucv.ro} \\
Faculty of Physics, University of Craiova\\
13 A. I. Cuza Str., Craiova 200585, Romania}
\title{No cross-interactions among different tensor fields with the mixed
symmetry $\left( 3,1\right) $ intermediated by a vector field}
\maketitle

\begin{abstract}
Under the hypotheses of analyticity in the coupling constant,
locality, Lorentz covariance, and Poincar\'{e} invariance of the
deformations, combined with the preservation of the number of
derivatives on each field, the consistent interactions between a
collection of free massless tensor gauge fields with the mixed
symmetry of a two-column Young diagram of the type (3,1) and one
Abelian vector field, respectively a $p$-form gauge field, are
addressed. The main result is that a single mixed symmetry tensor
field from the collection gets coupled to the vector field/$p$-form.
Our final result resembles to the well known fact from General
Relativity according to which there is one graviton in a given
world.

\end{abstract}

\section{Introduction}

Tensor fields in \textquotedblleft exotic\textquotedblright\ representations
of the Lorentz group, characterized by a mixed Young symmetry type~\cite%
{curt1,curt2,aul,labast1,labast2,burd,zinov1}, held the attention lately on
some important issues, like the dual formulation of field theories of spin
two or higher~\cite%
{dualsp1,dualsp2,dualsp2a,dualsp2b,dualsp3,dualsp4,dualsp5}, the
impossibility of consistent cross-interactions in the dual formulation of
linearized gravity~\cite{lingr}, a Lagrangian first-order approach~\cite%
{zinov2,zinov3} to some classes of massless or partially massive
mixed symmetry type tensor gauge fields, suggestively resembling to
the tetrad formalism of General Relativity, or the derivation of
some exotic gravitational interactions \cite{boulangerCQG,ancoPRD}.
An important matter related to mixed symmetry type tensor fields is
the study of their consistent interactions, among themselves as well
as with higher-spin gauge
theories~\cite{high1,high2,high3,high4,7,9,kk,3,4}. The most
efficient approach to this problem is the cohomological one, based
on the deformation of the solution to the master
equation~\cite{def}. The purpose of this paper is to investigate the
consistent interactions between a collection of massless tensor
gauge fields, each with the mixed symmetry of a two-column Young
diagram of the type $(3,1)$, and one vector field, respectively one
$p$-form gauge field. It is worth mentioning the duality of a free
massless tensor gauge field with the mixed symmetry $\left(
3,1\right) $ to the Pauli-Fierz theory in $D=6$ dimensions and, in
this respect, some developments concerning the dual formulations of
linearized gravity from the perspective of
$M$-theory~\cite{mth1,mth2,mth3}. Our analysis relies on the
deformation of the solution to the master equation by means of
cohomological techniques with the help of the local BRST cohomology,
whose component in a single $(3,1)$ sector has been reported in
detail in~\cite{t31jhep}. This paper generalizes our results
from~\cite{t31prd} regarding the cross-interactions between a single
massless $\left( 3,1\right) $ field and a vector field. Under the
hypotheses of analiticity in the coupling constant, locality,
Lorentz covariance, and Poincar\'{e} invariance of the deformations,
combined with the preservation of the number of derivatives on each
field, we find a deformation of the solution to the master equation
that provides nontrivial cross-couplings. This case corresponds to a
$p+4$-dimensional spacetime and is described by a deformed solution
that stops at order two in the coupling constant. The interacting
Lagrangian action contains only mixing-component terms of order one
and two in the coupling constant, \emph{but only one mixed symmetry
tensor field from the collection gets coupled to the $p$-form, while
the others remain free}. At the level of the gauge transformations,
only those of the $p$-form are modified at order one in the coupling
constant with a term linear in the antisymmetrized first-order
derivatives of a single gauge parameter from the $(3,1)$ sector such
that the gauge algebra and the reducibility structure of the coupled
model are not modified during the deformation procedure, being the
same like in the case of the starting free action. Our result is
interesting since it \emph{exhibits strong similarities to the
Einstein gravitons from General Relativity, in the sense that no
nontrivial cross-couplings between different fields with the mixed
symmetry }$\left( 3,1\right) $\emph{\ are allowed, neither direct
nor intermediated by a $p$-form}.

\section{Free model for $p=1$. BRST symmetry}

We begin with the Lagrangian action
\begin{eqnarray}
&&S_{0}\left[ t_{\lambda \mu \nu |\alpha }^{A},V_{\mu }\right] =\int
d^{D}x\left\{ \frac{1}{2}\left[ \left( \partial ^{\rho }t_{A}^{\lambda \mu
\nu |\alpha }\right) \left( \partial _{\rho }t_{\lambda \mu \nu |\alpha
}^{A}\right) -\left( \partial _{\alpha }t_{A}^{\lambda \mu \nu |\alpha
}\right) \left( \partial ^{\beta }t_{\lambda \mu \nu |\beta }^{A}\right) %
\right] \right.  \notag \\
&&-\frac{3}{2}\left[ \left( \partial _{\lambda }t_{A}^{\lambda \mu \nu
|\alpha }\right) \left( \partial ^{\rho }t_{\rho \mu \nu |\alpha
}^{A}\right) +\left( \partial ^{\rho }t_{A}^{\lambda \mu }\right) \left(
\partial _{\rho }t_{\lambda \mu }^{A}\right) \right] +3\left( \partial
_{\alpha }t_{A}^{\lambda \mu \nu |\alpha }\right) \left( \partial _{\lambda
}t_{\mu \nu }^{A}\right)  \notag \\
&&\left. +3\left( \partial _{\rho }t_{A}^{\rho \mu }\right) \left( \partial
^{\lambda }t_{\lambda \mu }^{A}\right) -\frac{1}{4}F_{\mu \nu }F^{\mu \nu
}\right\} \equiv S_{0}^{\mathrm{t}}\left[ t_{\lambda \mu \nu |\alpha }^{A}%
\right] +S_{0}^{\mathrm{V}}\left[ V_{\mu }\right] ,  \label{tv1}
\end{eqnarray}%
in $D\geq 5$ spacetime dimensions, with $A=\overline{1,n}$ and $n>1$. Each
massless tensor field $t_{\lambda \mu \nu |\alpha }^{A}$ has the mixed
symmetry $(3,1)$ and hence transforms according to an irreducible
representation of $GL(D,\mathbb{R})$ corresponding to a 4-cell Young diagram
with two columns and three rows. It is thus completely antisymmetric in its
first three indices and satisfies the identity $t_{[\lambda \mu \nu |\alpha
]}^{A}\equiv 0$. The collection indices $A$, $B$, etc., are raised and
lowered with a quadratic form $k_{AB}$ that defines a positively-defined
metric in the internal space. It can always be normalized to $\delta _{AB}$
by a simple linear field redefinition, so one can take $k_{AB}=\delta _{AB}$
and re-write (\ref{tv1}) as%
\begin{equation}
S_{0}\left[ t_{\lambda \mu \nu |\alpha }^{A},V_{\mu }\right] =\int d^{D}x%
\left[ \sum_{A=1}^{n}\mathcal{L}_{0}^{\mathrm{t}}\left( t_{\lambda \mu \nu
|\alpha }^{A},\partial _{\rho }t_{\lambda \mu \nu |\alpha }^{A}\right) +%
\mathcal{L}_{0}^{\mathrm{V}}\left( V_{\mu },\partial _{\nu }V_{\mu }\right) %
\right] ,  \label{tv1a}
\end{equation}%
where $\mathcal{L}_{0}^{\mathrm{t}}\left( t_{\lambda \mu \nu |\alpha
}^{A},\partial _{\rho }t_{\lambda \mu \nu |\alpha }^{A}\right) $ is the
Lagrangian density for the field $A$. The field strength of the vector field
$V_{\mu }$ is defined in the standard manner by
\begin{equation}
F_{\mu \nu }=\partial _{\mu }V_{\nu }-\partial _{\nu }V_{\mu }\equiv
\partial _{\left[ \mu \right. }V_{\left. \nu \right] }.  \label{abfstr}
\end{equation}%
Everywhere in this paper it is understood that the notation $[\lambda \cdots
\alpha ]$ signifies complete antisymmetry with respect to the (Lorentz)
indices between brackets, with the conventions that the minimum number of
terms is always used and the result is never divided by the number of terms.
The trace of $t_{\lambda \mu \nu |\alpha }^{A}$ is defined by $t_{\lambda
\mu }^{A}=\sigma ^{\nu \alpha }t_{\lambda \mu \nu |\alpha }^{A}$ and it is
obviously an antisymmetric tensor. Everywhere in this paper we employ the
flat Minkowski metric of `mostly plus' signature $\sigma ^{\mu \nu }=\sigma
_{\mu \nu }=(-,++++\cdots )$.

A generating set of gauge transformations for action (\ref{tv1}) can be
taken of the form
\begin{eqnarray}
\delta _{\epsilon ,\chi }t_{\lambda \mu \nu |\alpha }^{A} &=&-3\partial _{
\left[ \lambda \right. }\epsilon _{\left. \mu \nu \alpha \right]
}^{A}+4\partial _{\left[ \lambda \right. }\epsilon _{\left. \mu \nu \right]
\alpha }^{A}+\partial _{\left[ \lambda \right. }\chi _{\left. \mu \nu \right]
|\alpha }^{A},  \label{tv7a} \\
\delta _{\epsilon }V_{\mu } &=&\partial _{\mu }\epsilon ,  \label{tv7b}
\end{eqnarray}%
where the gauge parameters $\epsilon _{\lambda \mu \nu }^{A}$ determine $n$
completely antisymmetric tensors, the other set of gauge parameters displays
the mixed symmetry $\left( 2,1\right) $, such that each of them is
antisymmetric in the first two indices and satisfies the identity $\chi _{%
\left[ \mu \nu |\alpha \right] }^{A}\equiv 0$, and the gauge parameter $%
\epsilon $ is a scalar. The generating set of gauge transformations (\ref%
{tv7a})--(\ref{tv7b}) is off-shell, second-order reducible, the
accompanying gauge algebra being obviously Abelian (for details, see
\cite{t31jhep}).

The construction of the antifield-BRST symmetry for this free theory debuts
with the identification of the algebra on which the BRST differential $s$
acts. The generators of the BRST algebra are of two kinds: fields/ghosts and
antifields. The ghost spectrum for the model under study comprises the
fermionic ghosts $\left\{ \eta _{\lambda \mu \nu }^{A},\mathcal{G}_{\mu \nu
|\alpha }^{A},\eta \right\} $ associated with the gauge parameters $\left\{
\epsilon _{\lambda \mu \nu }^{A},\chi _{\mu \nu |\alpha }^{A},\epsilon
\right\} $ from (\ref{tv7a})--(\ref{tv7b}), the bosonic ghosts for ghosts $%
\left\{ C_{\mu \nu }^{A},\mathcal{C}_{\nu \alpha }^{A}\right\} $ due to the
first-stage reducibility relations, and also the fermionic ghosts for ghosts
for ghosts $C_{\nu }^{A}$ corresponding to the second-stage reducibility
relations. We ask that $\eta _{\lambda \mu \nu }^{A}$ and $C_{\mu \nu }^{A}$
are completely antisymmetric, $\mathcal{G}_{\mu \nu |\alpha }^{A}$ display
the mixed symmetry $\left( 2,1\right) $, and $\mathcal{C}_{\nu \alpha }^{A}$
are symmetric. The antifield spectrum is organized into the antifields $%
\left\{ t_{A}^{\ast \lambda \mu \nu |\alpha },V^{\ast \mu }\right\} $ of the
original tensor fields, together with those of the ghosts, $\left\{ \eta
_{A}^{\ast \lambda \mu \nu },\mathcal{G}_{A}^{\ast \mu \nu |\alpha },\eta
^{\ast }\right\} $, $\left\{ C_{A}^{\ast \mu \nu },\mathcal{C}_{A}^{\ast \nu
\alpha }\right\} $, and respectively $C_{A}^{\ast \nu }$, of statistics
opposite to that of the associated fields/ghosts. It is understood that $%
t_{A}^{\ast \lambda \mu \nu |\alpha }$ exhibit the same mixed-symmetry
properties like $t_{\lambda \mu \nu |\alpha }^{A}$ and similarly with
respect to $\eta _{A}^{\ast \lambda \mu \nu }$, $\mathcal{G}_{A}^{\ast \mu
\nu |\alpha }$, $C_{A}^{\ast \mu \nu }$, and $\mathcal{C}_{A}^{\ast \nu
\alpha }$. For subsequent purpose, we denote the trace of $t_{A}^{\ast
\lambda \mu \nu |\alpha }$ by $t_{A}^{\ast \lambda \mu }$, being understood
that it is antisymmetric.

Since both the gauge generators and reducibility functions for this model
are field-independent, it follows that the BRST differential $s$ simply
reduces to
\begin{equation}
s=\delta +\gamma ,  \label{tv39}
\end{equation}%
where $\delta $ represents the Koszul-Tate differential, graded by the
antighost number $\mathrm{agh}$ ($\mathrm{agh}\left( \delta \right) =-1$)
and $\gamma $ stands for the exterior derivative along the gauge orbits,
whose degree is named pure ghost number $\mathrm{pgh}$ ($\mathrm{pgh}\left(
\gamma \right) =1$). The overall degree that grades the BRST complex is
known as the ghost number ($\mathrm{gh}$) and is defined like the difference
between the pure ghost number and the antighost number, such that $\mathrm{gh%
}\left( s\right) =\mathrm{gh}\left( \delta \right) =\mathrm{gh}\left( \gamma
\right) =1$. According to the standard rules of the BRST method, the
corresponding degrees of the generators from the BRST complex are valued
like
\begin{equation*}
\mathrm{pgh}\left( \eta _{\lambda \mu \nu }^{A}\right) =\mathrm{pgh}\left(
\mathcal{G}_{\mu \nu |\alpha }^{A}\right) =\mathrm{pgh}\left( \eta \right)
=1,
\end{equation*}%
\begin{equation*}
\mathrm{pgh}\left( C_{\mu \nu }^{A}\right) =2=\mathrm{pgh}\left( \mathcal{C}%
_{\nu \alpha }^{A}\right) ,\quad \mathrm{pgh}\left( C_{\nu }^{A}\right) =3,
\end{equation*}%
\begin{equation*}
\mathrm{agh}\left( t_{A}^{\ast \lambda \mu \nu |\alpha }\right) =1=\mathrm{%
agh}\left( V^{\ast \mu }\right) ,
\end{equation*}%
\begin{equation*}
\mathrm{agh}\left( \eta _{A}^{\ast \lambda \mu \nu }\right) =\mathrm{agh}%
\left( \mathcal{G}_{A}^{\ast \mu \nu |\alpha }\right) =\mathrm{agh}\left(
\eta ^{\ast }\right) =2,
\end{equation*}%
\begin{equation*}
\mathrm{agh}\left( C_{A}^{\ast \mu \nu }\right) =3=\mathrm{agh}\left(
\mathcal{C}_{A}^{\ast \nu \alpha }\right) ,\quad \mathrm{agh}\left(
C_{A}^{\ast \nu }\right) =4,
\end{equation*}%
plus the usual rules that the degrees of the original fields, the antighost
number of the ghosts and the pure ghost number of the antifields all vanish.
The actions of $\delta $ and $\gamma $ on the generators from the BRST
complex are given by
\begin{equation}
\gamma t_{\lambda \mu \nu |\alpha }^{A}=-3\partial _{\left[ \lambda \right.
}\eta _{\left. \mu \nu \alpha \right] }^{A}+4\partial _{\left[ \lambda
\right. }\eta _{\left. \mu \nu \right] \alpha }^{A}+\partial _{\left[
\lambda \right. }\mathcal{G}_{\left. \mu \nu \right] |\alpha }^{A},\quad
\gamma V_{\mu }=\partial _{\mu }\eta ,  \label{tv49}
\end{equation}%
\begin{equation}
\gamma \eta _{\lambda \mu \nu }^{A}=-\frac{1}{2}\partial _{\left[ \lambda
\right. }C_{\left. \mu \nu \right] }^{A},\quad \gamma \eta =0,  \label{tv50}
\end{equation}%
\begin{equation}
\gamma \mathcal{G}_{\mu \nu |\alpha }^{A}=2\partial _{\left[ \mu \right.
}C_{\left. \nu \alpha \right] }^{A}-3\partial _{\left[ \mu \right.
}C_{\left. \nu \right] \alpha }^{A}+\partial _{\left[ \mu \right. }\mathcal{C%
}_{\left. \nu \right] \alpha }^{A},  \label{tv51}
\end{equation}%
\begin{equation}
\gamma C_{\mu \nu }^{A}=\partial _{\left[ \mu \right. }C_{\left. \nu \right]
}^{A},\quad \gamma \mathcal{C}_{\nu \alpha }^{A}=-3\partial _{\left( \nu
\right. }C_{\left. \alpha \right) }^{A},\quad \gamma C_{\nu }^{A}=0,
\label{tv52}
\end{equation}%
\begin{equation}
\gamma t_{A}^{\ast \lambda \mu \nu |\alpha }=\gamma V^{\ast \mu }=\gamma
\eta _{A}^{\ast \lambda \mu \nu }=\gamma \mathcal{G}_{A}^{\ast \mu \nu
|\alpha }=\gamma \eta ^{\ast }=0,  \label{tv53}
\end{equation}%
\begin{equation}
\gamma C_{A}^{\ast \mu \nu }=\gamma \mathcal{C}_{A}^{\ast \nu \alpha
}=\gamma C_{A}^{\ast \nu }=0,  \label{tv53a}
\end{equation}%
\begin{equation}
\delta t_{\lambda \mu \nu |\alpha }^{A}=\delta V_{\mu }=\delta \eta
_{\lambda \mu \nu }^{A}=\delta \mathcal{G}_{\mu \nu |\alpha }^{A}=\delta
\eta =0,  \label{tv54}
\end{equation}%
\begin{equation}
\delta C_{\mu \nu }^{A}=\delta \mathcal{C}_{\nu \alpha }^{A}=\delta C_{\nu
}^{A}=0,  \label{tv54a}
\end{equation}%
\begin{equation}
\delta t_{A}^{\ast \lambda \mu \nu |\alpha }=T_{A}^{\lambda \mu \nu |\alpha
},\quad \delta V^{\ast \mu }=-\partial _{\nu }F^{\nu \mu },\quad \delta \eta
_{A}^{\ast \lambda \mu \nu }=-4\partial _{\alpha }t_{A}^{\ast \lambda \mu
\nu |\alpha },  \label{tv55}
\end{equation}%
\begin{equation}
\delta \mathcal{G}_{A}^{\ast \mu \nu |\alpha }=-\partial _{\lambda }\left(
3t_{A}^{\ast \lambda \mu \nu |\alpha }-t_{A}^{\ast \mu \nu \alpha |\lambda
}\right) ,\quad \delta \eta ^{\ast }=-\partial _{\mu }V^{\ast \mu },
\label{tv56}
\end{equation}%
\begin{equation}
\delta C_{A}^{\ast \mu \nu }=3\partial _{\lambda }\left( \mathcal{G}%
_{A}^{\ast \mu \nu |\lambda }-\frac{1}{2}\eta _{A}^{\ast \lambda \mu \nu
}\right) ,\quad \delta \mathcal{C}_{A}^{\ast \nu \alpha }=\partial _{\mu }%
\mathcal{G}_{A}^{\ast \mu \left( \nu |\alpha \right) },  \label{tv57}
\end{equation}%
\begin{equation}
\delta C_{A}^{\ast \nu }=6\partial _{\mu }\left( \mathcal{C}_{A}^{\ast \mu
\nu }-\frac{1}{3}C_{A}^{\ast \mu \nu }\right) ,  \label{tv58}
\end{equation}%
where $T_{A}^{\lambda \mu \nu |\alpha }$ are minus the Euler-Lagrange
derivatives of action (\ref{tv1}) with respect to the field $t_{\lambda \mu
\nu |\alpha }^{A}$.

The Lagrangian BRST differential admits a canonical action in a structure
named antibracket and defined by decreeing the fields/ghosts conjugated with
the corresponding antifields, $s\cdot =\left( \cdot ,S\right) $, where $%
\left( ,\right) $ signifies the antibracket and $S$ denotes the canonical
generator of the BRST symmetry. It is a bosonic functional of ghost number
zero (involving both field/ghost and antifield spectra) that obeys the
master equation $\left( S,S\right) =0$. The master equation is equivalent
with the second-order nilpotency of $s$, where its solution $S$ encodes the
entire gauge structure of the associated theory. Taking into account the
formulas (\ref{tv49})--(\ref{tv58}) as well as the standard actions of $%
\delta $ and $\gamma $ in canonical form we find that the complete solution
to the master equation for the free model under study is given by
\begin{eqnarray}
S &=&S_{0}\left[ t_{\lambda \mu \nu |\alpha }^{A},V_{\mu }\right] +\int
d^{D}x\left[ t_{A}^{\ast \lambda \mu \nu |\alpha }\left( 3\partial _{\alpha
}\eta _{\lambda \mu \nu }^{A}+\partial _{\left[ \lambda \right. }\eta
_{\left. \mu \nu \right] \alpha }^{A}+\partial _{\left[ \lambda \right. }%
\mathcal{G}_{\left. \mu \nu \right] |\alpha }^{A}\right) \right.  \notag \\
&&-\frac{1}{2}\eta _{A}^{\ast \lambda \mu \nu }\partial _{\left[ \lambda
\right. }C_{\left. \mu \nu \right] }^{A}+\mathcal{G}_{A}^{\ast \mu \nu
|\alpha }\left( 2\partial _{\alpha }C_{\mu \nu }^{A}-\partial _{\left[ \mu
\right. }C_{\left. \nu \right] \alpha }^{A}+\partial _{\left[ \mu \right. }%
\mathcal{C}_{\left. \nu \right] \alpha }^{A}\right)  \notag \\
&&\left. +C_{A}^{\ast \mu \nu }\partial _{\left[ \mu \right. }C_{\left. \nu %
\right] }^{A}-3\mathcal{C}_{A}^{\ast \nu \alpha }\partial _{\left( \nu
\right. }C_{\left. \alpha \right) }^{A}+V^{\ast \mu }\partial _{\mu }\eta %
\right] \equiv S^{\mathrm{t}}+S^{\mathrm{V}}.  \label{tv60}
\end{eqnarray}

\section{Brief review of the deformation procedure}

There are three main types of consistent interactions that can be added to a
given gauge theory: The first type deforms only the Lagrangian action, but
not its gauge transformations. The second kind modifies both the action and
its transformations, but not the gauge algebra. The third, and certainly
most interesting category, changes everything, namely, the action, its gauge
symmetries, and the accompanying algebra.

The reformulation of the problem of consistent deformations of a given
action and of its gauge symmetries in the antifield-BRST setting is based on
the observation that if a deformation of the classical theory can be
consistently constructed, then the solution to the master equation for the
initial theory can be deformed into the solution of the master equation for
the interacting theory
\begin{equation}
\bar{S}=S+gS_{1}+g^{2}S_{2}+O\left( g^{3}\right) ,\quad \varepsilon \left(
\bar{S}\right) =0,\quad \mathrm{gh}\left( \bar{S}\right) =0,  \label{tv61}
\end{equation}%
such that
\begin{equation}
\left( \bar{S},\bar{S}\right) =0.  \label{tv62}
\end{equation}%
Here and in the sequel $\varepsilon \left( F\right) $ denotes the Grassmann
parity of $F$. The projection of (\ref{tv61}) on the various powers of the
coupling constant induces the following tower of equations:
\begin{eqnarray}
g^{0} &:&\left( S,S\right) =0,  \label{tv63} \\
g^{1} &:&\left( S_{1},S\right) =0,  \label{tv64} \\
g^{2} &:&\frac{1}{2}\left( S_{1},S_{1}\right) +\left( S_{2},S\right) =0,
\label{tv65} \\
&&\vdots  \notag
\end{eqnarray}%
The first equation is satisfied by hypothesis. The second governs the
first-order deformation of the solution to the master equation, $S_{1}$ and
shows that $S_{1}$ is a BRST co-cycle, $sS_{1}=0$. This means that $S_{1}$
pertains to the ghost number zero cohomological space of $s$, $H^{0}\left(
s\right) $, which is generically non-empty because it is isomorphic to the
space of physical observables of the free theory. The remaining equations
are responsible for the higher-order deformations of the solution to the
master equation. No obstructions arise in finding solutions to them as long
as no further restrictions, such as spacetime locality or Lorentz
covariance, are imposed. Obviously, only nontrivial first-order deformations
should be considered, since trivial ones ($S_{1}=sB$) lead to trivial
deformations of the initial theory and can be eliminated by convenient
redefinitions of the fields. Ignoring the trivial deformations, it follows
that $S_{1}$ is a nontrivial BRST-observable, $S_{1}\in H^{0}\left( s\right)
$. Once that the deformation equations (\ref{tv64})--(\ref{tv65}), etc.,
have been solved by means of specific cohomological techniques, from the
consistent nontrivial deformed solution to the master equation one can
extract all the information on the gauge structure of the resulting
interacting theory.

\section{Main results for $p=1$\label{mainres}}

The aim of this paper is to investigate the consistent interactions that can
be added to action (\ref{tv1}) without modifying either the field spectrum
or the number of independent gauge symmetries. This matter is addressed in
the context of the antifield-BRST deformation procedure described in the
above and relies on computing the solutions to the Eqs. (\ref{tv64})--(\ref%
{tv65}), etc., from the cohomology of the BRST differential. For obvious
reasons, we consider only analytic, local, and manifestly covariant
deformations and, meanwhile, restrict to Poincar\'{e}-invariant quantities,
i.e. we do not allow explicit dependence on the spacetime coordinates. The
analyticity of deformations refers to the fact that the deformed solution to
the master equation, (\ref{tv61}), can be expanded in a formal power series
in the coupling constant $g$ that makes sense and reduces to the original
solution (\ref{tv60}) in the free limit ($g=0$). Moreover, we ask that the
deformed gauge theory preserves the Cauchy order of the uncoupled model,
which enforces the requirement that the interacting Lagrangian is of maximum
order equal to two in the spacetime derivatives of the fields at each order
in the coupling constant. Here, we present the main results without
insisting on the cohomology tools required by the technique of consistent
deformations. The cohomological proofs are similar to those from~\cite%
{t31jhep} and ~\cite{t31prd} and will not be detailed in the sequel. There
appear two distinct solutions to (\ref{tv62}), which cannot coexist. This is
due to the the higher-order consistency equations of the deformation
procedure. More precisely, both types of solutions survive at the level of $%
S_{1}$, $S_{2}$, and $S_{3}$, but the existence of $S_{4}$ as solution to
the equation $\frac{1}{2}\left( S_{2},S_{2}\right) +\left(
S_{1},S_{3}\right) +\left( S_{4},S\right) =0$ is equivalent to the result
that they are mutually exclusive (for more details, see Appendix B from \cite%
{t31prd}).

The first type of deformed solution to the master equation (\ref{tv62}) that
is consistent to all orders in the coupling constant stops at order one in
the coupling constant and reads as%
\begin{equation}
\bar{S}=S+\frac{g}{3\cdot 4!}\int d^{5}x\,\varepsilon ^{\lambda \mu \nu \rho
\kappa }F_{\lambda \mu }F_{\nu \rho }V_{\kappa },  \label{tv117}
\end{equation}%
where $S$ is given in (\ref{tv60}) in $D=5$. It is important to stress that
this result is obstructed to higher dimensions, being the \emph{only}
possibility in $D\geq 5$ that complies with \emph{all} of our working
hypothese. Indeed, the Chern-Simons actions in $D>5$, $\int d
^{2k+1}x\,\varepsilon ^{\mu _{1} \mu _{2}\ldots \mu _{2k-1} \mu _{2k} \mu
_{2k+1}}F_{\mu _{1} \mu _{2}}\cdots F_{\mu _{2k-1} \mu _{2k}}V_{\mu _{2k+1}}$%
, with $k>2$, are ruled out by the derivative-order assumption since they
contain $k>2$ spacetime derivatives. The case described by (\ref{tv117}) is
not interesting since it provides no cross-couplings between the vector
field and the tensor field with the mixed-symmetry $\left( 3,1\right) $. It
simply restricts the free Lagrangian action (\ref{tv1}) to evolve on a
five-dimensional spacetime and adds to it a generalized Abelian Chern-Simons
term, without changing the original gauge transformations (\ref{tv7a})--(\ref%
{tv7b}) and, in consequence, neither the original Abelian gauge algebra nor
the reducibility structure.

The second type of full deformed solution to the master equation (\ref{tv62}%
) ends at order two in the coupling constant and is given by%
\begin{eqnarray}
\bar{S} &=&S+g\sum_{A=1}^{n}\left[ y^{A}\int d^{5}x\,\varepsilon ^{\lambda
\mu \nu \rho \kappa }\left( V_{\lambda }^{\ast }\mathcal{F}_{\mu \nu \rho
\kappa }^{A}-\frac{2}{3}F_{\lambda \mu }\partial _{\left[ \xi \right.
}t_{\left. \nu \rho \kappa \right] |\theta }^{A}\sigma ^{\theta \xi }\right) %
\right]   \notag \\
&&+\frac{16g^{2}}{3}\sum_{A,B=1}^{n}\left[ y^{A}y^{B}\int d^{5}x\left(
\partial _{\left[ \xi \right. }t_{\left. \nu \rho \kappa \right] |\theta
}^{A}\sigma ^{\theta \xi }\right) \partial ^{\left[ \xi ^{\prime }\right.
}t^{B\left. \nu \rho \kappa \right] |\theta ^{\prime }}\sigma _{\theta
^{\prime }\xi ^{\prime }}\right] ,  \label{tv119}
\end{eqnarray}%
where all $\mathcal{F}_{\mu \nu \rho \kappa }^{A}$ have the pure ghost
number equal to one and are defined like the antisymmetrized first-order
derivatives of the ghosts $\eta _{\nu \rho \kappa }^{A}$ from the sector $%
\left( 3,1\right) $%
\begin{equation}
\mathcal{F}_{\mu \nu \rho \kappa }^{A}\equiv \partial _{\lbrack \mu }\eta
_{\nu \rho \kappa ]}^{A}.  \label{ghostfstr}
\end{equation}%
These are in fact the only nontrivial elements with the pure ghost number
equal to one from the cohomology of the exterior derivative along the gauge
orbits, $H\left( \gamma \right) $. The quantities $y^{A}$ are $n$ arbitrary,
real numbers and $\varepsilon ^{\lambda \mu \nu \rho \kappa }$ is the
Levi-Civita symbol in $D=5$. We observe that this solution `lives' also in a
five-dimensional spacetime, just like the previous one. Of course, there
appears the natural question whether (\ref{tv119}) can be generalized to
higher dimensions. The answer is again negative (like with respect to (\ref%
{tv117})), but for quite different reasons. Without entering too many
details, we will expose here only the main argument for the existence of
these obstructions. If one analyzes separately the first-order deformation
of the solution to the master equation in the cross-interacting sector%
\footnote{%
meaning that we search only solutions $S_{1}$ to equation (\ref{tv64}) that
effectively couple BRST generators from the vector sector with those
belonging to the mixed symmetry sector}, then it can be shown (see Appendix
A of \cite{t31prd}, subsection 2 --- Computation of first-order
deformations) that $S_{1}$ ends non-trivially at antighost number one
\begin{equation}
S_{1}=\int d^{D}x\left( a_{0}+a_{1}\right) ,\qquad \mathrm{agh}\left(
a_{i}\right) =i,\qquad i=0,1,  \label{sup1}
\end{equation}%
with $a_{i}$ solutions to the equations
\begin{equation}
\gamma a_{1}=0,\qquad \delta a_{1}+\gamma a_{0}=\partial _{\mu }j_{0}^{\mu
},\qquad \mathrm{agh}\left( j_{0}^{\mu }\right) =0.  \label{sup2}
\end{equation}%
Looking at (\ref{tv50}), it is easy to see that the general form of the
(nontrivial) solution to the former equation from (\ref{sup2}) reads as
\begin{equation}
a_{1}=\sum_{A=1}^{n}\left( t^{\ast A\lambda \mu \nu |\alpha }M_{\lambda \mu
\nu \alpha }^{A}\eta +V_{\lambda }^{\ast }N^{A\lambda \mu \nu \rho \kappa }%
\mathcal{F}_{\mu \nu \rho \kappa }^{A}\right) ,  \label{sup3}
\end{equation}%
where $M_{\lambda \mu \nu \alpha }^{A}$ and $N^{A\lambda \mu \nu \rho \kappa
}$ are $\gamma $-closed quantities built out of the original fields (which
is the same with gauge-invariant elements since $\gamma $ acts on the
original fields through the gauge transformations modulo replacing the gauge
parameters with the ghosts) in order to ensure $\gamma a_{1}=0$. Since the
most general gauge-invariant quantities of the free model are the Abelian
field strength $F_{\mu \nu }$, the `curvature' tensors $K_{\lambda \mu \nu
\rho |\alpha \beta }^{A}\equiv \partial _{\alpha }\partial _{\lbrack \lambda
}t_{\mu \nu \rho ]|\beta }^{A}-\partial _{\beta }\partial _{\lbrack \lambda
}t_{\mu \nu \rho ]|\alpha }^{A}$, and their derivatives, it follows that the
tensors $M^{A}$ and $N^{A}$ appearing in (\ref{sup3}) are polynomials in $F$%
, $K^{A}$, and their subsequent derivatives (up to a finite order in order
to render local deformations). Imposing the derivative-order assumption, it
follows immediately that the functions of type $M^{A}$ are restricted to be
at most linear in $F$, while all $N^{A}$ must be constant (since otherwise
one infers interaction vertices with more than two spacetime derivatives).
Requiring the Lorentz covariance and Poincar\'{e} invariance, it follows
that the only possible candidates are:
\begin{equation}
M_{\lambda \mu \nu \alpha }^{A}=w^{A}F_{\lambda \mu }\sigma _{\nu \alpha
},\qquad N^{A\lambda \mu \nu \rho \kappa }=y^{A}\delta _{5}^{D}\varepsilon
^{\lambda \mu \nu \rho \kappa },  \label{sup4}
\end{equation}%
with $w^{A}$ and $y^{A}$ some arbitrary, real constants and $\delta
_{5}^{D}$ the Kronecker symbol. Replacing (\ref{sup4}) into
(\ref{sup3}) and acting with $\delta $ on the resulting expression,
it can be shown (see Appendix A of \cite{t31prd}, subsection 2 ---
Computation of first-order deformations) that the latter equation in
(\ref{sup2}) does not possess solutions with respect to $a_{0}$
unless
\begin{equation}
w^{A}=0,\qquad A=\overline{1,n}.  \label{sup5}
\end{equation}%
Inserting (\ref{sup5}) in (\ref{sup4}) and the corresponding functions in (%
\ref{sup3}), we find that the last component of $S_{1}$ takes the general
form
\begin{equation}
a_{1}=\delta _{5}^{D}\varepsilon ^{\lambda \mu \nu \rho \kappa
}V_{\lambda }^{\ast }\left[ \sum_{A=1}^{n}\left(
y^{A}\mathcal{F}_{\mu \nu \rho \kappa }^{A}\right) \right] ,
\label{sup6}
\end{equation}%
which is nothing but the first term from the sum in the right-hand of (\ref%
{tv119}) for $D=5$. Starting with this only possibility for $a_{1}$,
it is merely a matter of computation to show that the corresponding
deformed solution to the master equation, which is consistent to
\emph{all} orders in the coupling constant, is precisely
(\ref{tv119}). We can thus state that the source of obstructions to
generalizations of (\ref{tv119}) in higher dimesions ($D>5$) is
complex, being given by a combination of \emph{all} hypotheses:
locality, Lorentz covariance, Poincar\'{e} invariance, and
derivative-order assumption.

From (\ref{tv119}) we read all the information on the gauge structure of the
coupled theory. The terms of antighost number zero in (\ref{tv119}) provide
the Lagrangian action. They can be equivalently organized as%
\begin{equation}
\bar{S}_{0}\left[ t_{\lambda \mu \nu |\alpha }^{A},V_{\mu }\right] =S_{0}^{%
\mathrm{t}}\left[ t_{\lambda \mu \nu |\alpha }^{A}\right] -\frac{1}{4}\int
d^{5}x\,\bar{F}_{\mu \nu }\bar{F}^{\mu \nu },  \label{equivdeflag}
\end{equation}%
in terms of the deformed field strength
\begin{equation}
\bar{F}^{\mu \nu }=F^{\mu \nu }+\frac{4g}{3}\varepsilon ^{\mu \nu \alpha
\beta \gamma }\sum_{A=1}^{n}\left( y^{A}\partial _{\left[ \rho \right.
}t_{\left. \alpha \beta \gamma \right] |}^{A\;\;\;\;\;\;\;\;\;\rho }\right) ,
\label{deffstr}
\end{equation}%
where $S_{0}^{\mathrm{t}}\left[ t_{\lambda \mu \nu |\alpha }^{A}\right] $ is
the Lagrangian action of the massless tensor fields $t_{\lambda \mu \nu
|\alpha }^{A}$ appearing in (\ref{tv1}) in $D=5$. We observe that the action
(\ref{equivdeflag}) contains only mixing-component terms of order one and
two in the coupling constant. The piece of antighost number one appearing in
(\ref{tv119}) gives the deformed gauge transformations in the form%
\begin{eqnarray}
\bar{\delta}_{\epsilon ,\chi }t_{\lambda \mu \nu |\alpha }^{A} &=&-3\partial
_{\left[ \lambda \right. }\epsilon _{\left. \mu \nu \alpha \right]
}^{A}+4\partial _{\left[ \lambda \right. }\epsilon _{\left. \mu \nu \right]
\alpha }^{A}+\partial _{\left[ \lambda \right. }\chi _{\left. \mu \nu \right]
|\alpha }^{A},  \label{defgauge1} \\
\bar{\delta}_{\epsilon ,\chi }V^{\mu } &=&\partial ^{\mu }\epsilon
+4g\varepsilon ^{\mu \alpha \beta \gamma \delta }\sum_{A=1}^{n}\left(
y^{A}\partial _{\alpha }\epsilon _{\beta \gamma \delta }^{A}\right) .
\label{defgauge2}
\end{eqnarray}%
It is interesting to note that only the gauge transformations of the vector
field are modified during the deformation process. This is enforced at order
one in the coupling constant by a term linear in the antisymmetrized
first-order derivatives of some gauge parameters from the $(3,1)$ sector. At
antighost numbers strictly greater than one (\ref{tv119}) coincides with the
solution (\ref{tv60}) corresponding to the free theory. Consequently, the
gauge algebra and the reducibility structure of the coupled model are not
modified during the deformation procedure, being the same like in the case
of the starting free action (\ref{tv1}) with the gauge transformations (\ref%
{tv7a})--(\ref{tv7b}). It is easy to see from (\ref{equivdeflag}) and (\ref%
{defgauge1})--(\ref{defgauge2}) that if we impose the PT-invariance at the
level of the coupled model, then we obtain no interactions (we must set $g=0$%
\ in these formulas).

Action (\ref{equivdeflag}) seems to couple the vector field to each field $%
t_{\lambda \mu \nu |\alpha }^{A}$ (assuming all $y^{A}$ are nonvanishing)
and also to provide cross-couplings between different fields $t_{\lambda \mu
\nu |\alpha }^{A}$ (see the last term from the right-hand side of (\ref%
{tv119}) with $A\neq B$). We will show that it is in fact possible to
redefine both the fields $t_{\lambda \mu \nu |\alpha }^{A}$ and the
constants $y^{A}$ such that: 1. the vector field gets coupled to a single
mixed symmetry tensor field from the collection, and 2. the cross-couplings
between different fields $t_{\lambda \mu \nu |\alpha }^{A}$ are discarded.
In order to show this result, let us denote by $Y$ the matrix of elements $%
y^{A}y^{B}$. It is simple to see that the rank of $Y$ is equal to one. By an
orthogonal transformation $M$ we can always find a matrix $\hat{Y}$ of the
form%
\begin{equation}
\hat{Y}=M^{T}YM,  \label{w1}
\end{equation}%
with $M^{T}$ the transposed of $M$, such that $\hat{Y}$ is diagonalized and
a single diagonal element (for definiteness, we take the first) is
nonvanishing%
\begin{equation}
\hat{Y}^{11}=\sum_{A=1}^{n}\left( y^{A}\right) ^{2}\equiv y^{2},\qquad \hat{Y%
}^{1A^{\prime }}=\hat{Y}^{B^{\prime }1}=\hat{Y}^{A^{\prime }B^{\prime
}}=0,\qquad A^{\prime },B^{\prime }=\overline{2,n}.  \label{w2}
\end{equation}%
If we make the notation%
\begin{equation}
\hat{y}^{A}=M^{AC}y^{C},  \label{w3}
\end{equation}%
then relation (\ref{w2}) implies%
\begin{equation}
\hat{y}^{A}=y\delta _{1}^{A}.  \label{w4}
\end{equation}%
Now, we make the linear field redefinition%
\begin{equation}
t_{\lambda \mu \nu |\alpha }^{A}=M^{AC}\hat{t}_{\lambda \mu \nu |\alpha
}^{C},  \label{w5}
\end{equation}%
with $M^{AC}$ the elements of $M$. It is easy to see that this
transformation leaves $S_{0}^{\mathrm{t}}\left[ t_{\lambda \mu \nu |\alpha
}^{A}\right] $ invariant (it remains equal to a sum of free actions, one for
every transformed field $\hat{t}_{\lambda \mu \nu |\alpha }^{A}$ from the
collection) and, moreover, the deformed action (\ref{equivdeflag}) becomes%
\begin{equation}
\bar{S}_{0}\left[ t_{\lambda \mu \nu |\alpha }^{A},V_{\mu }\right] =S_{0}^{%
\mathrm{t}}\left[ \hat{t}_{\lambda \mu \nu |\alpha }^{A}\right] -\frac{1}{4}%
\int d^{5}x\,\bar{F}_{\mu \nu }^{\prime }\bar{F}^{\prime \mu \nu },
\label{w8}
\end{equation}%
where
\begin{equation}
\bar{F}^{\prime \mu \nu }=F^{\mu \nu }+\frac{4g}{3}y\varepsilon ^{\mu \nu
\alpha \beta \gamma }\partial _{\left[ \rho \right. }\hat{t}_{\left. \alpha
\beta \gamma \right] |}^{1\;\;\;\;\;\;\;\;\;\rho }.  \label{w9}
\end{equation}%
Action (\ref{w8}) is invariant under the gauge transformations%
\begin{eqnarray}
\delta _{\hat{\epsilon},\hat{\chi}}\hat{t}_{\lambda \mu \nu |\alpha }^{A}
&=&-3\partial _{\left[ \lambda \right. }\hat{\epsilon}_{\left. \mu \nu
\alpha \right] }^{A}+4\partial _{\left[ \lambda \right. }\hat{\epsilon}%
_{\left. \mu \nu \right] \alpha }^{A}+\partial _{\left[ \lambda \right. }%
\hat{\chi}_{\left. \mu \nu \right] |\alpha }^{A},  \label{w10} \\
\delta _{\hat{\epsilon},\hat{\chi}}V^{\mu } &=&\partial _{\mu }\epsilon
+4gy\varepsilon ^{\mu \alpha \beta \gamma \delta }\partial _{\alpha }\hat{%
\epsilon}_{\beta \gamma \delta }^{1},  \label{w10a}
\end{eqnarray}%
where the new gauge parameters are%
\begin{equation}
\hat{\epsilon}_{\mu \nu \alpha }^{A}=\epsilon _{\mu \nu \alpha
}^{B}M^{BA},\quad \hat{\chi}_{\mu \nu |\alpha }^{A}=\chi _{\mu \nu |\alpha
}^{B}M^{BA}.  \label{w11}
\end{equation}%
It is now clear that (\ref{w8}) decomposes into the action inferred in~\cite%
{t31prd} that couples only the first tensor field with the mixed symmetry $%
\left( 3,1\right) $ from the collection ($A=1$) to the vector field and a
sum of free actions for the remaining $\left( n-1\right) $ tensor fields
with the mixed symmetry $\left( 3,1\right) $. In conclusion, one cannot
couple different fields with the mixed symmetry $\left( 3,1\right) $ through
a vector field. A single field of this kind may be coupled nontrivially in $%
D=5$, while the others remain free.

It is important to stress that the problem of obtaining consistent
interactions depends strongly on the spacetime dimension. For instance, if
one starts with action (\ref{tv1}) in $D>5$, then one inexorably gets $\bar{S%
}=S$, so \emph{no} term can be added to either the original Lagrangian or
its gauge transformations.

\section{Generalization to an arbitrary $p$}

Although the main results discussed so far do not admit
generalizations to $D>5$ for a vector field, there exists a possible
generalization if one extends the form degree from one to an
arbitrary $p$. In this situation the starting point is given by a
free model describing a collection of $n$ massless tensor fields
$t_{\lambda \mu \nu |\alpha }^{A}$
and an Abelian $p$-form%
\begin{equation}
S_{0}\left[ t_{\lambda \mu \nu |\alpha }^{A},V_{\mu _{1}\ldots \mu _{p}}%
\right] =S_{0}^{\mathrm{t}}\left[ t_{\lambda \mu \nu |\alpha
}^{A}\right] +S_{0}^{\mathrm{V}}\left[ V_{\mu _{1}\ldots \mu
_{p}}\right] ,  \label{tp1}
\end{equation}%
where
\begin{equation}
S_{0}^{\mathrm{V}}\left[ V_{\mu _{1}\ldots \mu _{p}}\right] =-\frac{1}{%
2\cdot \left( p+1\right) !}\int d^{D}xF_{\mu _{1}\ldots \mu
_{p+1}}F^{\mu _{1}\ldots \mu _{p+1}}  \label{tp2}
\end{equation}%
and $S_{0}^{\mathrm{t}}\left[ t_{\lambda \mu \nu |\alpha
}^{A}\right] $ follows from formula (\ref{tv1}). The spacetime
dimension is subject to the
inequality%
\begin{equation}
D\geq \max \left( 5,p+1\right) ,  \label{D}
\end{equation}%
which ensures that the number of physical degrees of freedom of this
free model is nonnegative. The Abelian $p$-form field strength is
defined in
the usual manner as%
\begin{equation}
F_{\mu _{1}\ldots \mu _{p+1}}=\partial _{\lbrack \mu _{1}}V_{\mu
_{2}\ldots \mu _{p+1}]}.  \label{tp3}
\end{equation}%
Action (\ref{tp1}) is invariant under a generating set of gauge
transformations given by (\ref{tv7a}) for the fields $t_{\lambda \mu
\nu
|\alpha }^{A}$ and by%
\begin{equation}
\delta _{\overset{\left( 1\right) }{\rho }}V_{\mu _{1}\ldots \mu
_{p}}=\partial _{\left[ \mu _{1}\right. }\overset{\left( 1\right) }{\rho }%
_{\left. \mu _{2}\ldots \mu _{p}\right] }  \label{tp5}
\end{equation}%
for the Abelian $p$-form, where the gauge parameters
$\overset{\left( 1\right) }{\rho }_{\mu _{1}\ldots \mu _{p-1}}$ are
completely antisymmetric. The gauge symmetries of
$S_{0}^{\mathrm{t}}\left[ t_{\lambda \mu \nu |\alpha
}^{A}\right] $ are reducible of order two, while the gauge transformations (%
\ref{tp5}) are reducible of order $\left( p-1\right) $, such that
the overall reducibility order will be equal to $\max \left(
2,p-1\right) $.

The BRST algebra contains two types of generators: some from the
collection sector, described previously, and the others from the
$p$-form sector. The latter generators comprise the field $V_{\mu
_{1}\ldots \mu _{p}}$ and its
antifield $V_{\mu _{1}\ldots \mu _{p}}^{\ast }$, the ghosts $\left( \overset{%
\left( k\right) }{\xi }_{\mu _{1}\ldots \mu _{p-k}}\right) _{k=\overline{1,p}%
}$ corresponding to the gauge parameters ($k=1$) and to the
reducibility functions ($k=\overline{2,p}$), together with their
antifields  $\left( \overset{\left( k\right) }{\xi }_{\mu _{1}\ldots
\mu _{p-k}}^{\ast }\right) _{k=\overline{1,p}}$ (all these
generators define, where appropriate, antisymmetric tensors). The
solution to the master equation for this free model takes the simple
form
\begin{equation}
S=S^{\mathrm{t}}+S^{\mathrm{V}},  \label{sbardef}
\end{equation}%
where $S^{\mathrm{t}}$ follows from (\ref{tv60}) and
$S^{\mathrm{V}}$ is
expressed by%
\begin{eqnarray}
S^{\mathrm{V}} &=&S_{0}^{\mathrm{V}}\left[ V_{\mu _{1}\ldots \mu
_{p}}\right] +\int d^{D}x\left( V^{\ast \mu _{1}\ldots \mu
_{p}}\partial _{\left[ \mu
_{1}\right. }\overset{\left( 1\right) }{\xi }_{\left. \mu _{2}\ldots \mu _{p}%
\right] }\right.   \notag \\
&&\left. +\sum_{k=1}^{p-1}\overset{\left( k\right) }{\xi }^{\ast \mu
_{1}\ldots \mu _{p-k}}\partial _{\left[ \mu _{1}\right.
}\overset{\left( k+1\right) }{\xi }_{\left. \mu _{2}\ldots \mu
_{p-k}\right] }\right) . \label{tp16}
\end{eqnarray}

Although the cohomological structure in the case of a $p$-form with
$p>1$ is clearly richer than in the presence of a vector field,
nevertheless the cohomology of the tensor fields with the mixed
symmetry $\left( 3,1\right) $ is dominant. Just like in the previous
situation of a vector field, there
appear two types of fully deformed solutions to the master equation, \emph{%
which again cannot coexist}. We cannot stress enough that these
results take place for the same working hypotheses like in the case
of a one-form. The
first type generalizes (\ref{tv117}) and is expressed by%
\begin{eqnarray}
\bar{S} &=&S+gc_{1}\delta _{2p+1}^{D}\int d^{D}x\varepsilon
^{\lambda _{1}\ldots \lambda _{p}\mu _{1}\ldots \mu
_{p+1}}V_{\lambda _{1}\ldots
\lambda _{p}}F_{\mu _{1}\ldots \mu _{p+1}} \nonumber \\
&&+gc_{2}\delta _{3p+2}^{D}\int d^{D}x\varepsilon ^{\lambda
_{1}\ldots \lambda _{p}\mu _{1}\ldots \mu _{p+1}\nu _{1}\ldots \nu
_{p+1}}V_{\lambda _{1}\ldots \lambda _{p}}F_{\mu _{1}\ldots \mu
_{p+1}}F_{\nu _{1}\ldots \nu _{p+1}} \label{chern},
\end{eqnarray}%
where $S$ reads as in (\ref{sbardef}), $c_{1,2}$ are two arbitrary,
real constants, and $\delta _{2p+1}^{D}$ denotes the Kronecker
symbol. This situation describes no interactions among the tensor
fields $t_{\lambda \mu
\nu |\alpha }^{A}$ or between $t_{\lambda \mu \nu |\alpha }^{A}$ and the $p$%
-form: it simply adds to the original Lagrangian density two
Chern-Simons terms (only for $p$ odd, since otherwise they are
trivial), without modifying the original gauge symmetries. The only
difference from the vector field case is that here two kinds of
Chern-Simons terms with at most two spacetime derivatives are
admitted (again, for an odd $p$), while there only one was allowed.
This is purely a matter of spacetime dimension since here $2p+1>\max
\left( 5,p+1\right) $ for any odd $p>1$, while for $p=1$ we have
that $2p+1=3<\max \left( 5,p+1\right) =5$. The second case is pictured by%
\begin{eqnarray}
\bar{S} &=&S+g\sum_{A=1}^{n}\left[ y^{A}\int d^{p+4}x\left(
\varepsilon ^{\lambda _{1}\ldots \lambda _{p}\mu \nu \rho \kappa
}V_{\lambda _{1}\ldots \lambda _{p}}^{\ast }\mathcal{F}_{\mu \nu
\rho \kappa }^{A}\right. \right.
\notag \\
&&\left. \left. +\left( -\right) ^{p}\frac{4}{3\cdot \left( p+1\right) !}%
\varepsilon ^{\lambda _{1}\ldots \lambda _{p+1}\nu \rho \kappa
}F_{\lambda _{1}\ldots \lambda _{p+1}}\partial _{\left[ \xi \right.
}t_{\left. \nu \rho
\kappa \right] |\theta }^{A}\sigma ^{\theta \xi }\right) \right]   \notag \\
&&+\frac{16g^{2}}{3}\sum_{A,B=1}^{n}\left[ y^{A}y^{B}\int
d^{p+4}x\left(
\partial _{\left[ \xi \right. }t_{\left. \nu \rho \kappa \right] |\theta
}^{A}\sigma ^{\theta \xi }\right) \partial ^{\left[ \xi ^{\prime
}\right. }t^{B\left. \nu \rho \kappa \right] |\theta ^{\prime
}}\sigma _{\theta ^{\prime }\xi ^{\prime }}\right]   \label{tpn86}
\end{eqnarray}%
and generalizes the result (\ref{tv119}). (It is clear that in the
limit $p=1 $ formula (\ref{tpn86}) is nothing but (\ref{tv119}).) It
describes a theory in $D=p+4$ spacetime dimensions that is valid for
\emph{any} value (even or odd) $p>1$, which couples the tensor
fields $t_{\mu \nu \lambda |\rho }^{A}$ to the $p$-form. Regarding
the Lagrangian structure of this coupled model, we mention that the
terms of antighost number zero present in (\ref{tpn86})
produce the Lagrangian action%
\begin{equation}
\bar{S}_{0}\left[ t_{\lambda \mu \nu |\alpha }^{A},V_{\lambda
_{1}\ldots \lambda _{p}}\right] =S_{0}^{\mathrm{t}}\left[ t_{\lambda
\mu \nu |\alpha }^{A}\right] -\frac{1}{2\cdot \left( p+1\right)
!}\int d^{p+4}x\bar{F}_{\mu _{1}\ldots \mu _{p+1}}\bar{F}^{\mu
_{1}\ldots \mu _{p+1}},  \label{tp87}
\end{equation}%
in terms of the deformed field strength
\begin{equation}
\bar{F}^{\mu _{1}\ldots \mu _{p+1}}=F^{\mu _{1}\ldots \mu
_{p+1}}+\left( -\right) ^{p+1}\frac{4g}{3}\varepsilon ^{\mu
_{1}\ldots \mu _{p+1}\alpha \beta \gamma }\sum_{A=1}^{n}\left(
y^{A}\partial _{\left[ \rho \right. }t_{\left. \alpha \beta \gamma
\right] |}^{A\;\;\;\;\;\;\;\;\;\rho }\right) , \label{tp88}
\end{equation}%
where $S_{0}^{\mathrm{t}}\left[ t_{\lambda \mu \nu |\alpha
}^{A}\right] $ is the free action for the collection of $\left(
3,1\right) $ mixed-symmetry type tensor fields evolving on a
spacetime of dimension $D=p+4$. The pieces of antighost number one
from (\ref{tpn86}) emphasize the deformed gauge
transformations (\ref{defgauge1}) in $D=p+4$ and%
\begin{equation}
\bar{\delta}_{\epsilon ,\chi }V^{\lambda _{1}\ldots \lambda
_{p}}=\partial ^{\mu }\epsilon +4g\varepsilon ^{\lambda _{1}\ldots
\lambda _{p}\mu \nu \rho \kappa }\sum_{A=1}^{n}\left( y^{A}\partial
_{\alpha }\epsilon _{\beta \gamma \delta }^{A}\right) ,
\label{tp89b}
\end{equation}%
such that only the gauge symmetries of the $p$-form are modified. If in (\ref%
{tp87}) and its gauge symmetries we perform the transformations (\ref{w4}), (%
\ref{w5}), and (\ref{w11}), then the cross-couplings between
different tensor fields $t_{\lambda \mu \nu |\alpha }^{A}$
intermediated by a $p$-form get decoupled and we are led to the same
conclusions like in the case of a vector field: \emph{the
}$p$\emph{-form interacts with a single tensor field
(}$\hat{t}_{\lambda \mu \nu |\alpha }^{1}$\emph{), while the remaining }$%
\left( n-1\right) $\emph{\ tensor fields with the mixed symmetry
}$\left( 3,1\right) $\emph{\ are left free}.

\section{Conclusions}

The main conclusion of this paper is the proof of rigidity of the
couplings of a collection of tensor fields with the mixed symmetry
$\left( 3,1\right) $ to a vector field and actually to an arbitrary
$p$-form gauge field. This means that under some natural assumptions
(analyticity of the
deformations in the coupling constant, locality, Lorentz covariance, Poincar%
\'{e} invariance, and preservation of the number of derivatives on
each field), a single mixed symmetry tensor field from the
collection gets coupled to the vector field (or to a $p$-form). Our
final result resembles to the well known fact from General
Relativity according to which there is one graviton in a given
world. This is not a surprise since the action of a free tensor
field with the mixed symmetry $\left( 3,1\right) $ is dual to the
linearized gravity (in $D=6$).

\section*{Acknowledgment}

The authors are partially supported by the European Commission FP6 program
MRTN-CT-2004-005104 and by the contract CEx 05-D11-49/7.10.2005 with ANCS
(Romanian National Authority for Scientific Research).

\end{document}